\documentclass[aps,prl,reprint,superscriptaddress,nofootinbib,twocolumn,hidelinks]{revtex4-2}

\usepackage{graphicx}
\usepackage{rotating}
\usepackage{dcolumn}
\usepackage{amssymb}
\usepackage{amsmath}
\usepackage{bm}
\usepackage{times}
\usepackage{listings}
\usepackage{fancyhdr}
\usepackage{wrapfig}
\usepackage[caption=false]{subfig}
\usepackage{xcolor}
\usepackage{hyperref}
\hypersetup{
    colorlinks,
    linkcolor={red!50!black},
    citecolor={blue!50!black},
    urlcolor={blue!80!black}
}

\newcommand{\Dquad}{\qquad\qquad}
\newcommand{\nnn}{\nonumber \\}

\newcommand{\beeq}{\begin{equation}}
\newcommand{\eneq}{\end{equation}}
\newcommand{\bear}{\begin{eqnarray}}
\newcommand{\enar}{\end{eqnarray}}

\newcommand{\rbar}{\bar r}  
     
\newcommand{\HH}{\mathcal{H}} 
\newcommand{\OO}{\mathcal{O}}
\newcommand{\al}{\alpha}
\newcommand{\be}{\beta}
\newcommand{\ga}{\gamma}
\newcommand{\de}{\delta}
\newcommand{\para}{\parallel}
\newcommand{\pa}{\partial}
\newcommand{\bobs}{\bar{\rm o}}
\newcommand{\RR}{\mathcal{R}}

\allowdisplaybreaks

\begin{document}

\title{Conditions for the Absence of Infrared Sensitivity in Cosmological Probes in Any Gravity Theories}

\author{Matteo Magi}
\email[]{matteo.magi@uzh.ch}
\affiliation{Center for Theoretical Astrophysics and Cosmology, Institute for Computational Science, University of Zurich, CH--8057 Zurich, Switzerland}

\author{Jaiyul Yoo}
\email[]{jyoo@physik.uzh.ch}
\affiliation{Center for Theoretical Astrophysics and Cosmology, Institute for Computational Science, University of Zurich, CH--8057 Zurich, Switzerland}
\affiliation{Physics Institute, University of Zurich, Winterthurerstrasse 190, CH--8057, Zurich, Switzerland}

\date{\today}

\begin{abstract}
Large-scale surveys allow us to construct cosmological probes such as galaxy clustering, weak gravitational lensing, the luminosity distance, and cosmic microwave background anisotropies. The gauge-invariant descriptions of these cosmological probes reveal the presence of numerous relativistic effects in the cosmological probes, and they are sensitive (or even divergent) to the long wave-length fluctuations in the initial conditions. In the standard $\Lambda$CDM model, this infrared sensitivity is absent due to subtle cancellations among the relativistic contributions, once the Einstein equation is used. Here we derive the most general conditions for the absence of infrared sensitivity in the cosmological probes without committing to general relativity. We discuss the implications of our results for gravity theories beyond general relativity.
\end{abstract}

\maketitle

{\it Introduction.}---
Cosmological observations in large-scale surveys
allow us to construct cosmological probes such as galaxy clustering, weak gravitational lensing, the luminosity distance, and cosmic microwave background (CMB) anisotropies with ever increasing precision (see, e.g., \cite{LSST04,SKA09,EUCLID11,WFIRST12,DESI13,SPHER14} for the upcoming surveys). In particular, the two-point statistics of these cosmological probes are often used to constrain the cosmological models (see, e.g., \cite{PEBAET01,EIZEET05,TEEIET06,KODUET08,PLANCKcos18}). These two-point (and also $N$-point) statistics contain critical information about the initial conditions and the evolution of fluctuations, which are set by inflation and evolved according to general relativity in the standard cosmological model. For instance, galaxy clustering in a simple description measures the matter density fluctuations at the redshift of a galaxy sample, and CMB anisotropies are a snapshot of the gravitational potential fluctuations at the recombination epoch.

However, these cosmological probes are obtained by measuring light signals from a source such as galaxies, CMB photons, and it involves light propagation from the source to the observer, which are affected by relativistic effects such as the gravitational redshift, the gravitational lensing effect and so on \cite{SAWO67,KAISE92}. Hence, all the cosmological probes are in fact affected by subtle relativistic effects, which in turn provide a great opportunity to probe the early universe and gravity on large scales with the relativistic effects 
\cite{SASAK87,FUSA89,BODUGA06,YOFIZA09,YOO10,CHLE11,BODU11,JESCHI12,BEDUET14,UMCLMA14a,YOZA14,BEMACL14,DIDUET14,BOCLET15a,KOUMET18,MAYO22},
but at the same time it can result in sensitivity to the presence of long wave-length fluctuations or infrared sensitivity in the cosmological probes.

Consider the two-point correlation function of a cosmological probe~${\cal P}$
in a given survey, and the long wave-length contributions to the two-point
correlation is
	\bear\label{div}
	\langle {\cal P}(\eta,\bm x){\cal P}(\eta,\bm y) \rangle \simeq
\int d\ln k~{\cal T}^2_{\cal P}(\eta,k)\Delta^2_{\cal R}(k)~,
	\enar
where ${\cal T}_{\cal P}$ is the transfer function and
$\Delta^2_{\cal R}:=A_s (k/k_0)^{n_s-1}$ is the scale-invariant primordial power spectrum with scalar spectral index $n_s\simeq1$ and amplitude $A_s\simeq10^{-9}$ at a given scale~$k_0$ \cite{PLANCKcos15,PLANCKcos18}. 
The presence of gravitational potential in the cosmological probe~${\cal P}$ implies an enhanced sensitivity to long wave-length fluctuations: the integral is dominated by the contributions at very low Fourier modes~$k$ and may become logarithmically divergent in the infrared limit $k\to0$, as the transfer function
for gravitational potential on large scales is a constant.

Any cosmological probes are modulated by long wave-length fluctuations (see, e.g., \cite{HUKR03}), but their sensitivity is expected to gradually wane as the wave-length increases beyond the characteristic scale, i.e., the distance between the observer and the sources. In the case of gravitational potentials, such sensitivity would amount to a violation of the equivalence principle, since the leading contributions of long wave-length fluctuations correspond to a spatially uniform gravitational field. In fact, it has been shown in the past that this infrared sensitivity is absent for galaxy clustering \cite{JESCHI12,SCYOBI18,GRSCET20,CADI22}, luminosity distance \cite{BIYO16,BIYO17}, and CMB temperature anisotropies \cite{BAYO21} at the linear order in perturbations by assuming a $\Lambda$CDM model and solving the Einstein equations in general relativity. 

However, these cosmological probes or their theoretical descriptions are not limited to the underlying model of matter contents or gravity theories like those in $\Lambda$CDM models adopted in the previous work. The only assumptions behind the theoretical descriptions of the cosmological probes are geodesic equation for photons and the Friedmann-Lema\^itre-Robertson-Walker (FLRW) metric. So it is interesting to ask ``is the absence of infrared sensitivity specific to a $\Lambda$CDM model with general relativity? What happens to the cosmological probes in terms of infrared sensitivity in gravity theories beyond general relativity?''
In this Letter, we investigate the infrared sensitivity of the cosmological probes without assuming any gravity model, and we derive generic conditions that any gravity theories should satisfy to ensure the absence of infrared sensitivity.

\textit{Cosmological probes.}--- Here we consider theoretical descriptions for galaxy clustering, the luminosity distance, and the CMB temperature anisotropies and investigate their sensitivity to long wave-length fluctuations. These three cosmological probes are extensively used in large-scale surveys to constrain cosmological models. Though weak gravitational lensing is also a powerful cosmological probe (e.g., \cite{MELLI99,BASC01,REFRE03,HEAVE03,MUVAET08}), the relativistic contributions to the weak lensing signals come from 
at least two spatial derivatives of gravitational potential,
and hence it is rather insensitive to long wave-length fluctuations of our interest here.

The cosmological probes are constructed in terms of basic observables, and in cosmology these observables are the observed redshift (or spectral information), the angular position of the light sources, and the number of the observed sources or the observed flux in case the sources are not spatially resolved. Under the minimal assumption that the light propagation follows null geodesics, these observables can then be related to the physical quantities of the sources by tracing back the null path from the observer position to the source position \cite{YOO09,YOFIZA09,YOO14a,YOGRET18}. Since the observables are measured in our rest frame and the physical quantities of the sources are defined in the rest frame of the sources, their relation connected via a null path is completely independent of the coordinate system or diffeomorphism invariant \cite{YODU17,MIYO20}. While this description of the cosmological observables involves the matter contents in terms of the observer and the source, the relation between the matter and the light propagation (or metric fluctuations) is left free, such that its validity is {\it not} limited to general relativity.

The theoretical description of the luminosity distance is essentially the physical area of the source in the rest frame, corresponding to the observed angular size at the observed redshift \cite{SASAK87,BODUGA06,BODUKU06,BEGAET12a,BEDUET14,YOSC16,BIYO16,BIYO17},
which defines the luminosity distance (or the angular diameter distance) in case of the standard candle, for which the luminosity in the rest frame is known. Similar to the case for the luminosity distance, the theoretical description of galaxy clustering is the physical volume of the source in the rest frame, with the additional dimension corresponding to the observed redshift interval 
\cite{YOFIZA09,YOO10,CHLE11,BODU11,JESCHI12,BRCRET12,UMCLMA14a,YOZA14,
DIDUET14,BEMACL14,MAJOET20}. 
Furthermore, galaxy clustering contains an intrinsic fluctuation in relation to the matter density fluctuations, as different galaxy samples cluster differently (or known as galaxy bias \cite{PRSC74,KAISE84,POWI84,BBKS86,BCEK91}). Finally, the theoretical description of CMB temperature anisotropies is essentially the same as the observed redshift, under the assumption that the recombination takes place instantaneously at the temperature set by the atomic physics \cite{ZISC08,YOMIET19T,YOMIET19,BAYO21}, since its ratio to the observed temperature in the sky is the (observed) redshift.

{\it Linear-order expressions}--- Here we adopt the spatially flat FLRW metric for our cosmological model. In the Newtonian gauge the metric reads
\bear\label{Newton}
ds^2=-a^2(1 + 2\psi)d\eta^2+a^2[( 1+2\phi)\de_{\al\be}+2h_{\al\be}]dx^\al dx^\be,
\enar
where $a(\eta)$ is the scale factor as a function of conformal time, $\psi(\eta,\bm x)$ and $\phi(\eta,\bm x)$ are the Newtonian-gauge potentials while $h_{\al\be}(\eta,\bm x)$ is the transverse-traceless tensor perturbation. Here we perform only linear-order computations and ignore the vector contributions. Any time-like motion in this spacetime is described by a four-velocity $u^\mu=a^{-1}(1-\psi,-\pa^\al v)^\mu$, where the peculiar velocity potential $v$ arises from the inhomogeneities of the universe. We assume that the observer and light sources move along geodesics but we will relax this assumption later.

Now we present the linear-order expressions for the cosmological observables.
The expression for the luminosity distance fluctuation in the Newtonian gauge is given by \cite{YOSC16,BIYO16,BIYO17,MAYO22} as
	\bear\label{D}
	\de D_L=\de z+\frac{\de r}{\rbar_z}-\kappa+\phi -\frac12h_{\para\para}\,,
	\enar
and the expression for the observed galaxy number density fluctuation is given by \cite{YOFIZA09,SCYOBI18,GRSCET20,MAYO22} as
	\bear\label{gal}
	\de_g=b\left(\de_m+3\HH v_m\right)-e_z\left(\de z+\HH v_m\right)+\de V\,,
	\enar
where the distortion $\de V$ in the volume occupied by the source is
	\bear\label{V}
	\de V=2~\de D_L+\de z+\phi-\pa_\para v+h_{\para\para}+\frac{d}{d\rbar}\de r\,,
	\enar
$\HH=a'/a$ is the conformal Hubble function, $\de z$ is the perturbation in the observed redshift, $\de r$ is the radial distortion in the observed source position at comoving distance $\rbar_z$, $\kappa$ is the gravitational lensing convergence, $\de_m=\de\rho_m/\bar\rho_m$ is the matter density fluctuation, $b$ is the galaxy bias, $e_z$ is the evolution bias parameter of the mean galaxy number density. We defined the line-of-sight derivative
	\bear\label{los}
	\frac{d}{d\rbar}=n^\al\pa_\al-\pa_\eta=:\pa_\para-\pa_\eta\,,
	\enar
and the subscript $\para$ denotes the contraction of a spatial index with the observation direction $n^\al$. Finally, the expression for the observed CMB temperature anisotropies is \cite{YOMIET19,YOMIET19T,BAYO21}
	\bear\label{CMB}
	\Theta=\left(\frac14\de_\ga-\de z\right)_*\,,
	\enar
where $\de_\ga=\de\rho_\ga/\bar\rho_\ga$ is the radiation density fluctuation and all the terms are evaluated at the recombination position indicated by $*$. For the derivations and detailed expressions of the redshift perturbation $\de z$, lensing convergence $\kappa$, and radial distortion $\de r$, we refer the reader to the aforementioned references (see, e.g., \cite{YOGRET18}). The structure of those quantities at the linear order is rather simple, consisting of contributions at the observer position, source position, and integrations along the line-of-sight direction.

{\it Infrared contributions in the cosmological probes}--- 
The observer-source system in cosmology has a finite extent defined by the comoving distance $\rbar_z$ out to the observed redshift of the light source. Since we want to address the long wave-length behavior of the perturbations, consider a comoving scale $R$ much larger than $\rbar_z$ (ideally larger than any sound horizon of the involved propagating fields). Denoting any scalar or tensor perturbation as $f$, we define the long wave-length part $f_L$ of the perturbation $f$ by smoothing it over the scale~$R$ (the exact form of the smoothing kernel is irrelevant here). The leftover of this procedure is the short wave-length contribution $f_S:=f-f_L$, which starts with at least $k^2 f$ in Fourier space for a Gaussian smoothing kernel and hence will not be considered anymore. By definition, the long-mode $f_L$ varies very little in space over the scale $R$, thus the successive terms in the following spatial expansion around an arbitrary origin
	\bear\label{IR}
	f_L(\eta,\bm x)=:f_L(\eta)+x^\al\pa_\al f_L(\eta)+\OO(\bm x^2)\,,
	\enar
can be made arbitrarily small according to the choice of $R$. The perturbations of our interest are those on the line-of-sight of the observer and hence their spatial dependence can be parameterized by the comoving distance and angles, more conveniently in terms of
	\bear\label{exp}
	f_0(\eta):=f_L(\eta)\,,\Dquad f_1(\eta,\bm n):=\pa_\para f_L(\eta)\,,
	\enar
where the subscripts $_0$ and $_1$ represent the zeroth and first-order terms in the spatial expansion. The spatial expansion of $f_L$ in Eq.\,($\ref{IR}$) corresponds in Fourier space to 
	\bear\label{Fourier}
	f_L(\eta,\bm x)=\int \frac{d^3k}{(2\pi)^3}\left[ 1+i \bm k\cdot\bm x+\OO(\bm k^2) \right]f_L(\eta,\bm k)\,,~~~~
	\enar
such that higher-order terms in the spatial expansion are multiplied by higher powers of $k$ and the infrared behaviors are hence controlled by the lowest-order terms $f_0,f_1$ in the expansion. As evident in Eq.\,($\ref{div}$) these infrared modes in the cosmological probes would lead to either infrared divergences or sensitivity of the cosmological probes to super-horizon modes especially given the scale-invariant fluctuations as the initial conditions.

To study the infrared contributions in the cosmological probes we can apply the spatial expansion to all the perturbation variables in Eqs.\,($\ref{D}$)$-$($\ref{CMB}$) and keep only the lowest-order terms. Such procedure is straightforward but results in lengthy expressions. To simplify such expressions, we first introduce the comoving gauge ($v\equiv0$) curvature perturbation $\RR$ and the uniform-density gauge ($\de\rho\equiv0$) curvature perturbation $\zeta$:
	\bear\label{curv}
	\RR:=\phi-\HH v\,,\qquad
	\zeta:=\phi-\HH\frac{\de\rho}{\bar\rho'}\,,
	\enar
here the peculiar velocity $v$ and the energy density $\rho:=\bar\rho+\de\rho$ are the total ones, i.e.,
	\bear\label{phi}
	v=\frac{ 3\bar\rho_m v_m+4\bar\rho_\ga v_\ga }{3\bar\rho_m+4\bar\rho_\ga}\,,\qquad \de\rho=\de\rho_m+\de\rho_\ga\,,
	\enar
for models with matter ($m$) and radiation ($\ga$) fluids.
Second, we exploit the covariant conservation of the energy-momentum tensor, which we assume to be separately conserved for multiple non-interacting fluids such as matter and photon fluids today. Mind that the conservation equations hold for any gravity theory formulated in the Jordan frame. The continuity equation for linear perturbations yields
	\bear\label{continuity0}
	\de'_m+3\phi'=\Delta v_m\,,\qquad\de'_\ga+4\phi'=\Delta v_\ga\,,
	\enar
where the Laplacians can be neglected in the infrared, leading to the following expressions
	\bear\label{zete}
	\zeta'_m=\zeta'_\ga=0\,,\qquad \zeta_\ga-\zeta_m=\frac{\de_\ga}4-\frac{\de_m}3\,,
	\enar
obtained using the definition in Eq.\,($\ref{curv}$) for each fluid. For the spatial components of the conservation equations we derive
	\bear\label{continuityiforte}
	\psi=\HH v_m+v'_m\,,\qquad \psi= v'_\ga-\frac{\de p_\ga}{4\bar p_\ga}
	\,,
	\enar
where $\de p_\ga$ is the radiation pressure fluctuation around the background $\bar p_\ga$.
Utilizing Eq.\,($\ref{los}$) to replace the line-of-sight integration 
in the infrared with the boundary terms at the observer ($\eta_{\bobs}$)
and source position ($\eta_z$)
	\bear
	\int_0^{\rbar_z}d\rbar\,\pa_\eta f_{0,1}(\eta)=f_{0,1}(\eta_{\bobs})-f_{0,1}(\eta_z)\,,
	\enar
and combining all previous equations, we obtain the zeroth-order terms of the cosmological probes in the spatial expansion:
	\bear\label{prima}
	\de D_{L,0}&=&\left(\frac1{\rbar_z\HH_z}-1\right)\left[\RR_m+h_{\para\para}\right]^z_{\bobs}
		- \frac32h_{\para\para}(\eta_{\bobs})
	\nnn
	&&
	-\frac12h_{\para\para}
	+\RR_m
	-\frac1{\rbar_z}\int_0^{\rbar_z} d\rbar\,\left(\RR_m-2h_{\para\para}\right)\,,\qquad
	\\
	\nnn
	\de_{g,0}&=&3b\left(\zeta_m-\RR_m\right)+e_z\left[\RR_m+h_{\para\para}\right]^z_{\bobs}+\de V_0\,,
	\\
	\nnn
	\de V_0&=&2\de D_{L,0}-\left(1-\frac{\HH'_z}{\HH^2_z} \right)\left[\RR_m+h_{\para\para}\right]^z_{\bobs}
	-\frac{\RR'_m+h'_{\para\para}}{\HH_z}\,,
	\nnn
	\\
	\nnn
	\Theta_0&=&\zeta_\ga-\RR_m(\eta_{\bobs})+h_{\para\para}-h_{\para\para}(\eta_{\bobs})\,,
	\enar
with all the variables in the expressions representing the zeroth-order terms $f_0(\eta)$ in the spatial expansion, and the first-order terms
	\bear
	\de D_{L,1}&=&-\left(\frac{1}{\rbar^2_z\HH_z}\right)\int_0^{\rbar_z} d\rbar\left(\RR_m+h_{\para\para}\right)-\frac32h_{\para\para}
	\nnn&&
	+\frac{\RR_m+h_{\para\para}}{\rbar_z\HH_z}+\int_0^{\rbar_z} d\rbar\,\left(\frac{5\rbar-\rbar_z}{\rbar^2_z}\right)h_{\para\para}\,,\qquad
	\\
	\nnn
	\de_{g,1}&=&\de V_1+e_z\bigg(\RR_m+h_{\para\para}
	-\frac1{\rbar_z}\int_0^{\rbar_z} d\rbar\left(\RR_m+h_{\para\para}\right)\bigg)\nnn&& +3b\left(\zeta_m-\RR_m\right)
	\,,
	\\
	\nnn
	\de V_1&=&2~\de D_{L,1}-\frac{\RR'_m+h'_{\para\para}}{\HH_z}
	-\left(1-\frac{\HH'_z}{\HH^2_z} \right)\times
	\nnn
	&&
	\bigg(\RR_m+h_{\para\para}-\frac1{\rbar_z}\int_0^{\rbar_z} d\rbar\,\left(\RR_m+h_{\para\para}\right)\bigg)\,,
	\\
	\nnn
	\label{ultima}
	\Theta_1&=&
	\zeta_\ga+h_{\para\para}+\frac{v_\ga-v_m}{\rbar_*}-\frac1{\rbar_*}\int_0^{\rbar_*} d\rbar\,\left(\RR_m+h_{\para\para}\right),
	\nnn
	\enar
with all the perturbation variables representing the first-order terms $f_1(\eta,\bm n)$ in the spatial expansion.
We emphasize that these expressions are the leading infrared contributions in the cosmological probes, for example, ${\cal P}_L(\eta,\bm x)={\cal P}_0(\eta)+\rbar_z{\cal P}_1(\eta,\bm n)+\OO(\bm x^2)$, and they are obtained without committing to any gravity theory. These equations are the main results of the current work.

{\it General conditions for the absence of infrared sensitivity.}---
With
a closer look to Eqs.\,($\ref{prima}$)$-$($\ref{ultima}$), it is possible to derive necessary and sufficient conditions for the probes to be devoid of infrared contributions or the lowest-order terms ${\cal P}_0$ and ${\cal P}_1$. First of all, the tensor perturbations must be constant in time in the leading spatial expansion:
	\bear\label{tensor}
	\left(h'_{\al\be}\right)_{0,1}=0\,,
	\enar
for all the cosmological probes.
For the luminosity distance it is further necessary that the comoving-gauge curvature perturbation is time independent in the infrared
	\bear\label{Rm}
	\left(\RR'_m\right)_{0,1}=0\,.
	\enar
For galaxy clustering, the presence of galaxy bias in proportion to the matter density fluctuation in the proper-time hypersurface imposes an extra condition in addition to Eqs.\,($\ref{tensor}$) and ($\ref{Rm}$):
	\bear\label{Rzeta}
	\RR_{m,0}=\zeta_{m,0}\,\Dquad \RR_{m,1}=\zeta_{m,1}
	\,,
	\enar
i.e., the comoving-gauge curvature perturbations cannot be any arbitrary constant in the infrared, but it needs to be equal to the uniform-density gauge curvature perturbation. In the case of the CMB anisotropies, since both matter and radiation fluids are involved as the observer and the source, the conditions in the scalar sector now read
	\bear\label{cmb}
	v_{m,1}(\eta_*)=v_{\ga,1}(\eta_*)\,,\quad
	\RR_{m,0}=\zeta_{\ga,0}\,,\quad \RR_{m,1}=\zeta_{\ga,1}\,,\quad
	\enar
in addition to Eqs.\,($\ref{tensor}$) and ($\ref{Rm}$).
We can combine all the conditions stated so far, such that any cosmological probes presented here are devoid of infrared contributions, if and only if
	\beeq\label{condit0}
	\begin{split}
	\left(\RR'_m\right)_{0,1}&=\left(h'_{\al\be}\right)_{0,1}=0\,,\qquad v_{m,1}(\eta_*)=v_{\ga,1}(\eta_*)\,,
	\\
	\RR_{m,0}&=\zeta_{m,0}=\zeta_{\ga,0}\,,\qquad\RR_{m,1}=\zeta_{m,1}=\zeta_{\ga,1}\,.
	\end{split}
	\eneq
We can recast these equations in terms of fluid quantities via Eqs.\,($\ref{curv}$) and ($\ref{zete}$):
	\beeq\label{condit}
	\begin{split}
	\left(\frac{\de\rho_{m}}{\bar\rho'_m}-\frac{\de\rho_{\ga}}{\bar\rho'_\ga}\right)_{0,1}=0=\left( \frac{\de\rho_m}{\bar\rho'_m}-v_m \right)_{0,1}\,,
	\\
	 v_{m,1}(\eta_*)=v_{\ga,1}(\eta_*)\,,\Dquad 	\left(h'_{\al\be}\right)_{0,1}=0\,.
	 \end{split}
	\eneq
These conditions are the adiabatic conditions for the fluids 
and the conservation of $\RR$ and $h_{\al\be}$ in the infrared. Note, however,
that the equality condition between the velocity and the density is in general
determined not only by the properties of the fluids, but also by the 
underlying gravity. We stress again that the derivation of 
Eq.\,($\ref{condit}$) was made without resorting to $\Lambda$CDM or
 general relativity.

{\it Discussion on the general conditions.}--- The theoretical descriptions
of the cosmological probes and hence the general conditions are derived
by assuming that photons travel on a null geodesic and the matter components
follow a time-like geodesic. The matter components (the observer or the source)
can, of course, deviate from a geodesic motion, if they experience 
non-gravitational forces. In general, these forces fall off sufficiently fast
on large scales. To be precise, if the two lowest-order terms of the
non-gravitational forces vanish in the spatial expansion,  
our derivation of the general conditions is not affected.

Infrared divergences or very large fluctuations in the cosmological probes
would arise, if the general conditions are broken at all~$k$ on large scales.
If they are broken at some~$k$, but not at all~$k$, no infrared divergences
exist, but the cosmological probes are still sensitive to and modulated
by infrared modes. Such infrared sensitivity is not as strong, 
as the divergence is only logarithmic, if $n_s\simeq1$ on such large scales.
However, the sensitivity can be further enhanced, if the spectral index~$n_s$
becomes redder ($n_s\ll1$) on larger scales. 

Critical to our conclusions is the existence of fluctuations on all large 
scales. A presence of a cut-off scale~$k_{\rm IR}$ in the infrared, beyond which
there is {\it no} fluctuation at all, would simply eliminate the infrared
divergences or limit the sensitivity in the cosmological probes. 
Such cut-off scales might naturally
arise from the trans-Planckian conjecture \cite{BRMA13,BEBRET20,BLLABR23}.
While an
inflationary expansion in the early Universe lasts for a finite number of
$e$-folding, it is somewhat odd to demand that the Universe is perfectly
homogeneous and isotropic without any fluctuations, even at the level 
of $10^{-5}$ on scales larger than the largest scale with fluctuations
generated by inflation. Nonetheless, the status of such large scales is
governed by pre-inflationary physics and is highly uncertain.

{\it Implications for cosmological models beyond the standard.}---
In the standard $\Lambda$CDM model with general relativity, the adiabatic 
conditions on large scales are \textit{assumed} for the matter and radiation components 
presumably arising from the decay of the inflaton field. With the adiabatic
 conditions, the conservation of~$\zeta$ in the infrared is guaranteed by 
the energy momentum conservation, while the infrared conservation of~$\RR$ 
is guaranteed only by the Einstein equations. The equivalence of~$\RR$ 
and~$\zeta$ in the infrared is, however, imposed only by the Einstein
 equations, independent of the adiabatic conditions for the matter components. 
For the infrared 
conservation of the tensor fluctuations $h_{\al\be}$, the Einstein equations 
are needed in conjunction with the absence of anisotropic pressure 
in the infrared.
It was shown \cite{MIYOMA23} that this infrared insensitivity
in the cosmological probes holds non-linearly in $\Lambda$CDM models.

Our general conditions apply to any  four-dimensional metric theory of gravity with diffeomorphism symmetry. 
In such modified gravity theories, there exists at least one additional degree
of freedom (see, e.g., \cite{DETS10,CLFEET12} for recent reviews),
arising from higher-order derivatives in the equation of
motion.  Despite these extra degrees of freedom, adiabatic conditions 
for matter and radiation components can be assumed in the same way as 
in the $\Lambda$CDM model, and the conservation of the energy-momentum 
tensor still leads to the conservation of~$\zeta$ in the infrared. 
The conservation of~$\RR$ in the infrared is, on the other hand,
 dictated by the modified Einstein equations and may be allowed.
Unlike in general relativity, conservation of both~$\RR$ and~$\zeta$ may 
not result in their equivalence. Consequently, our general conditions can 
constrain the space of physically viable solutions in any gravity theories.

Furthermore, with extra degrees of freedom, the adiabaticity alone in the
matter sector would not be enough to meet the general conditions, as the
extra degrees of freedom will act as non-adiabatic fluctuations on large 
scales, though a complete synchronization might be still possible.
We note, however, that
any viable gravity theories often admit a solution that mimics
general relativity in some limit such as the Brans-Dicke model \cite{BRDI61}
and the Horndeski theories \cite{HORND74}.
If such solution exists on large scales,
we expect the general conditions to be met, but of course such solution
is {\it not} the most interesting one in any modified gravity theories. 

Multi-field inflationary models
\cite{MOLLE90,POST92,BAMARI01,BYWA06,GOWAET00,PETE11}
or models with cosmological defects
(see, e.g., \cite{BRAND94,BRAND14} for review)
often generate isocurvature perturbations. The presence of isocurvature
fluctuations on very large scales would break the general conditions and lead to
the infrared sensitivity. Long-range non-gravitational forces 
would act in a similar way to break the general conditions, and
the cosmological probes would be modulated by very long wave-length modes.
These forces can arise from the presence of primordial non-Gaussianity
in the initial condition \cite{KOSP01,BAKOET04},
from an interaction in the dark sector \cite{ARCAET22}, or
cosmological back-reactions
\cite{BUEH97,ABBRMU97,BUCHE00,GEBR02,CLELET11,MAVABR13,COBR23}.
However, the level of these non-adiabatic fluctuations
is already tightly constrained by CMB observations
\cite{PLANCKfnl19,PLANCKinf20}, and hence 
the infrared sensitivity from these models is expected to be negligible.

We acknowledge useful discussions with 
Robert Brandenberger, Ermis Mitsou, and Marko Simonovi{\'c}.
This work is support by the Swiss National Science Foundation
and a Consolidator Grant of the European Research Council.

\bibliography{ms.bbl}

\end{document}